\documentclass[11pt, twocolumn]{article}
\usepackage[margin=0.8in]{geometry}
\usepackage[utf8]{inputenc}

\usepackage{amsmath,amssymb,amsfonts,amsthm,color,soul}

\usepackage{graphicx}
\usepackage{subfigure}

\usepackage[usenames]{xcolor}

\usepackage{bm}

\usepackage{url}
\usepackage{hyperref}
\hypersetup{
  colorlinks   = true, 
  urlcolor     = blue, 
  linkcolor    = black, 
  citecolor   = black 
}

\usepackage{lipsum}

\usepackage{calc}

\newcommand\blfootnote[1]{
  \begingroup
  \renewcommand\thefootnote{}\footnote{#1}
  \addtocounter{footnote}{-1}
  \endgroup
}

\usepackage[numbers,sort&compress]{natbib}
\let\OLDthebibliography\thebibliography
\renewcommand\thebibliography[1]{
  \OLDthebibliography{#1}
  \setlength{\parskip}{2pt}
  \setlength{\itemsep}{0pt plus 0.3ex}
}

\allowdisplaybreaks

\newcommand{\gev}{~\mathrm{GeV}}
\newcommand{\tev}{~\mathrm{TeV}}

\newcommand\refeq[1]{Eq.~(\ref{#1})}

\newcommand\refta[1]{Tab.~\ref{#1}}
\newcommand\refse[1]{Sect.~\ref{#1}}

\newcommand\citere[1]{Ref.~\cite{#1}}
\newcommand\citeres[1]{Refs.~\cite{#1}}

\def\reffi#1{\mbox{Fig.~\ref{#1}}}

\newcommand{\muCMS}{0.33}
\newcommand{\dmuCMSpl}{0.19}
\newcommand{\dmuCMSmi}{0.12}
\newcommand{\sigATLAS}{\SH{1.7}}
\newcommand{\muATLAS}{\SH{0.18}}
\newcommand{\dmuATLASpl}{\SH{0.10}}
\newcommand{\dmuATLASmi}{\SH{0.10}}
\newcommand{\muACgaga}{\mu_{\gamma\gamma}^{{\rm ATLAS+CMS}}}
\newcommand{\muAgaga}{\mu_{\gamma\gamma}^{{\rm ATLAS}}}
\newcommand{\muCgaga}{\mu_{\gamma\gamma}^{{\rm CMS}}}
\newcommand{\muAC}{\SH{0.24}}
\newcommand{\dmuACp}{\SH{0.09}}
\newcommand{\dmuACm}{\SH{0.08}}
\newcommand{\sigAC}{\SH{3.1}}

\newcommand{\massATLAS}{\SH{95.4}}
\newcommand{\massAC}{\SH{95.4}}
\newcommand{\simmass}{\SH{95}}

\newcommand{\hnf}{\ensuremath{h_{95}}}
\newcommand{\hotf}{\ensuremath{h_{125}}}

\newcommand\plane[2]{$(#1, #2)$ plane}

\newcommand{\SH}[1]{{\color{black}#1}}
\newcommand{\hto}[1]{{\color{black}#1}}

\newcommand{\TB}[1]{{\color{black}#1}}
\newcommand{\TBn}[1]{{\color{black}#1}}

\newcommand{\GW}[1]{{\color{black}#1}}
\newcommand{\GWn}[1]{{\color{black}#1}}
\newcommand{\GWnn}[1]{{\color{black}#1}}

\newcommand{\htr}[1]{{\color{black} #1}}
\newcommand{\TBB}[1]{{\color{black}#1}}

\usepackage{abstract}

\author{thomas.biekoetter@kit.edu}

\date{\today}

\usepackage{titlesec}
\titleformat{\section}
{\normalfont\bfseries}{\thesection}{1em}{}
\titleformat{\subsection}
{\normalfont\bfseries}{\thesubsection}{1em}{  }

\begin{document}

\def\thefootnote{\fnsymbol{footnote}}

\twocolumn[
\begin{@twocolumnfalse}
\begin{flushright}
\footnotesize
  KA-TP-11-2023, ~~ 
  DESY-23-071, ~~ 
  IFT--UAM/CSIC-23-062 ~~
\end{flushright}

\begin{center}
{\large
\textbf{
The 95.4 GeV di-photon excess at ATLAS and CMS 
}
}
\vspace{0.4cm}

Thomas Biekötter$^1$\footnote{thomas.biekoetter@desy.de},
Sven Heinemeyer$^2$\footnote{Sven.Heinemeyer@cern.ch} and
Georg Weiglein$^{3,4}$\footnote{georg.weiglein@desy.de}\\[0.2em]

{\small

  $^1${\textit{
   Institute for Theoretical Physics,
   Karlsruhe Institute of Technology,\\
   Wolfgang-Gaede-Str.~1, 76131 Karlsruhe, Germany
 }}
 
 $^2${\textit{
   Instituto de Física Teórica UAM-CSIC, Cantoblanco, 28049,
   Madrid, Spain
 }}
 
 $^3${\textit{
   Deutsches Elektronen-Synchrotron DESY,
     Notkestr.~85, 22607 Hamburg, Germany
  }}
  
 $^4${\textit{
   II. Institut für Theoretische Physik, Universität Hamburg,\\
    Luruper Chaussee 149, 22761 Hamburg, Germany\\[0.4em]
  }}
  
}

\begin{abstract}
The ATLAS collaboration has recently reported the
results of a low-mass Higgs-boson search in the di-photon 
final state based on the full Run 2 data set. 
The results are based on an improved analysis w.r.t.~the previous analysis, which included a part of the 
Run~2 data, with a substantially 
better sensitivity. The ``model-dependent'' search
carried out by ATLAS shows an excess
of events at a mass of about
$\massATLAS \gev$ with a local significance of
$\sigATLAS\,\sigma$.
The results are 
compatible with a previously 
reported excess at the 
same mass, but somewhat higher
significance of $2.9\,\sigma$,
from the CMS collaboration, also based on the 
full Run~2 data set. Combining the two 
results (neglecting possible correlations)
we find a signal strength
of $\muACgaga = \muAC^{+\dmuACp}_{-\dmuACm}$,
corresponding to an excess of $\sigAC\,\sigma$.
In this work, we 
investigate the implications of this result,
updating a previous analysis based solely 
on the CMS Run~2 data. We demonstrate that the ATLAS/CMS
combined di-photon excess
can be interpreted as the
lightest Higgs boson in a
Two-Higgs doublet
model that is extended by a complex singlet (S2HDM) of
Yukawa types~II and~IV,
while being in agreement with all other experimental
and theoretical constraints.
\end{abstract}

\end{center}

\end{@twocolumnfalse}
]

\renewcommand{\thefootnote}{\arabic{footnote}}
\setcounter{footnote}{0}

\section{Introduction}
\label{sec:intro}

More than a decade after
the discovery of a Higgs boson with a mass of 
about $125 \gev$ by the ATLAS and CMS 
collaborations~\cite{Aad:2012tfa,Chatrchyan:2012xdj},
the search for additional Higgs bosons 
continues to be one of the 
prime tasks of the LHC physics program.
\blfootnote{\hspace{-6mm}$^*$\mbox{}thomas.biekoetter@kit.edu,
$^\dagger$sven.heinemeyer@cern.ch,\hfill\mbox{}\\
$^\ddagger$georg.weiglein@desy.de
}
Searches for Higgs bosons below
$125\gev$ have been performed at
LEP~\cite{Abbiendi:2002qp,Barate:2003sz,
Schael:2006cr},
the Tevatron~\cite{Group:2012zca} and the
LHC~\cite{CMS:2015ocq,Sirunyan:2018aui,
CMS:2018rmh,ATLAS:2018xad,CMS:2022goy,ATLAS:2022abz,
CMSnew,ATLAS-CONF-2023-035}.
Among them, 
CMS has performed searches for scalar di-photon
resonances at~$8\tev$ and $13\tev$.
Results based on the $8\tev$ data and the
full Run~2 data set at $13\tev$
showed a local excess of $2.9\,\sigma$
at $95.4 \gev$~\cite{CMSnew}.
This excess, which is present in both the $8\tev$
and the $13\tev$ data set, 
received considerable attention 
already soon after it was made public using the 
first year of Run~2 data, 
see e.g.~\citeres{Cao:2016uwt,
Fox:2017uwr,
Richard:2017kot,
Haisch:2017gql,
Biekotter:2017xmf,
Liu:2018xsw,
Domingo:2018uim,
Biekotter:2019kde,
Cline:2019okt,
Cao:2019ofo,
Aguilar-Saavedra:2020wrj}. 
Analyses using the result based on the
full Run~2 data
can  be found in \citere{Biekotter:2023jld,
Azevedo:2023zkg} 
(see also \citere{Biekotter:2023qbe}). 

Recently, ATLAS presented the result based on
their full Run~2 data set~\cite{ATLASnewtalk,
ATLAS-CONF-2023-035}.
\htr{In the following, we refer to the ``model-dependent''
analysis from \mbox{ATLAS}, 
which has a higher discriminating power.}
The new analysis has a 
substantially improved sensitivity
w.r.t.~their analysis based on 
the previously reported result
utilizing $80\,\mathrm{fb}^{-1}$~\cite{ATLAS:2018xad}.
ATLAS finds an excess with a local
significance of $\sigATLAS\,\sigma$ at 
\GW{precisely the same mass value as the one that was previously reported by CMS, namely at}
$\massATLAS \gev$.
This ``di-photon excess'' can be
described by a scalar
resonance at $\massATLAS \gev$ with a signal strength
of 
\begin{equation}
\muAgaga =\frac{\sigma^{\rm exp} \left( \TBn{pp} \to \phi \to \gamma\gamma \right)}
         {\sigma^{\rm SM}\left( \TBn{pp} \to H \to \gamma\gamma \right)}
     = \muATLAS^{+\dmuATLASpl}_{-\dmuATLASmi} \ ,
\label{muATLAS}
\end{equation}
which we determined based on the reported
expected and observed limits at $\massATLAS \gev$
and the reported significance of the excess.
\TBn{Here $\sigma^{\rm SM}$ denotes the cross section
for a hypothetical SM
Higgs boson at the same mass.
Since ATLAS presented their results
as limits on the total cross section, we normalized
these limits with the SM prediction
$\sigma^{\rm SM}(pp \to H \to \gamma\gamma)
= 126~\mathrm{pb}$~\cite{LHCHiggsCrossSectionWorkingGroup:2016ypw}
in order to find the value for
$\muAgaga$ shown in \refeq{muATLAS}.}
The corresponding CMS result for a mass
of $95.4 \gev$ is given by~\cite{CMSnew}
\begin{equation}
\muCgaga =\frac{\sigma^{\rm exp} \left( \TBn{pp} \to \phi \to \gamma\gamma \right)}
         {\sigma^{\rm SM}\left( \TBn{pp} \to H \to \gamma\gamma \right)}
     = \muCMS^{+\dmuCMSpl}_{-\dmuCMSmi} \ .
\label{muCMS}
\end{equation}

Regarding the interpretation of the new result from ATLAS
together with the previously reported one from CMS, 
it is important to note
that a possible signal at about $95\gev$
giving rise to a relatively small number of events 
would occur 
on top of a much larger
fluctuating background. Therefore, 
one cannot necessarily expect that the excesses should 
occur \GW{with exactly the same signal strength, 
and the fact that both collaborations report their
most significant excess at precisely the same
mass value has to be seen in this context
as a certain level of coincidence.} 
\GW{Since for the same mass value the signals strengths 
$\muAgaga$ and $\muCgaga$ agree with each other within their 
uncertainties,}
we regard the two 
\GW{results}
to be 
compatible with each other. 
\GW{It should also be noted in this context, see \reffi{fig:gamgam} 
below, that the upper bound observed by ATLAS at $\massAC \gev$, albeit slightly stronger than the one observed by CMS at this mass value, lies significantly above the signal interpretation of the CMS result that is reflected in $\muCgaga$.}
Neglecting possible 
correlations we obtain a combined signal strength of
\begin{equation}
    \mu^\mathrm{exp}_{\gamma\gamma} = \muACgaga = \muAC^{+\dmuACp}_{-\dmuACm}\,,
    \label{muAC}
\end{equation}
corresponding to an excess of $\sigAC\,\sigma$ 
\GW{at}
\begin{equation}
  m_{\phi} \equiv m_{\phi}^{\rm ATLAS+CMS} = \massAC \gev\,.
\label{massAC}
\end{equation}

If the origin of the di-photon excesses
at $\massAC \gev$ is a new particle,
which is the scenario that we investigate here,
the question arises whether it is also detectable in
other collider channels.
\TB{In addition, the new particle could
have been produced already in small numbers
in other existing searches.
In this regard, it is interesting to note that}
LEP reported a local $2.3\,\sigma$ excess
in the~$e^+e^-\to Z(\phi\to b\bar{b})$
searches\,\cite{Barate:2003sz}, which would
be consistent with a
scalar resonance with a mass of about 
$\massAC \gev$
and a signal strength of
$\mu_{bb}^{\rm exp} =
0.117 \pm 0.057$~\cite{Cao:2016uwt,Azatov:2012bz}.
In addition to the di-photon excess,
CMS observed another excess
compatible with a mass of $\massAC \gev$ in
the Higgs-boson searches utilizing
di-tau final states~\cite{CMS:2022goy}.
This excess was
most pronounced at a mass of $100\gev$
with a local significance of $3.1\,\sigma$,
but it is also well compatible with a mass
of $\massAC \gev$, where the local significance
amounts to $2.6\,\sigma$,
and where the corresponding
signal strength for a mass hypothesis
of $95 \gev$ 
was determined to be
$\mu^{\rm exp}_{\tau\tau} =
1.2 \pm 0.5$.
ATLAS has not yet published a search in
the di-tau final state that covers the
mass range around 95~GeV.

Given that all the excesses discussed above
occurred at a similar mass,
\TB{it is possible that they}
arise from the 
production of a single new particle -- 
which would be a
first sign of physics beyond the SM (BSM)
in the Higgs-boson sector.
This triggered activities in the literature regarding possible 
model interpretations that could account 
for the various excesses~\cite{Cao:2016uwt,
Fox:2017uwr,
Richard:2017kot,
Haisch:2017gql,
Biekotter:2017xmf,
Liu:2018xsw,
Domingo:2018uim,
Biekotter:2019kde,
Cline:2019okt,
Cao:2019ofo,
Aguilar-Saavedra:2020wrj,
Biekotter:2021ovi,
Biekotter:2021qbc,Heinemeyer:2021msz,
Biekotter:2022jyr,Biekotter:2022abc,
Iguro:2022dok,
Biekotter:2023jld,
Azevedo:2023zkg}.
The first analysis using the CMS result based on the full Run~2 data
can  be found in \citere{Biekotter:2023jld}.

\GW{Since the new result from ATLAS implies that a moderate}
di-photon excess at about
95~GeV 
\GW{has independently been} observed by
two different experiments,
\GW{it is of interest to assess
the implications
of the combined result from ATLAS and CMS} on possible
model interpretations.
In the present paper we focus in particular on 
the extension of the 2HDM by a complex singlet
(S2HDM) as a template for a model where 
a mostly gauge-singlet scalar particle
obtains its couplings to fermions
and gauge bosons via the mixing with
the SM-like Higgs boson at $125\gev$.
We will demonstrate that 
this kind of scenario is suitable for describing
the di-photon excess, taking into account the 
\GW{(in comparison to the CMS result slightly increased)}
significance 
\GW{of the combined result.}
Moreover, we will discuss the possibility
of simultaneously
describing the $b\bar b$ excess
and the di-tau excess.

The paper is structured as follows. In \refse{sec:modeldef} 
we briefly introduce the S2HDM and 
define our notation.
In \refse{sec:quanti} 
we provide a brief
qualitative discussion on how
sizable signal rates
in the three channels in which the excesses
have been observed can arise.
The relevant theoretical and experimental
constraints on the model parameters
are briefly summarized
in \refse{sec:constraints}.
We present our main results regarding the 
numerical analysis of the
improved significance
of the di-photon excess in
\refse{sec:num}.
The conclusions and an outlook are given
in \refse{sec:conclu}.

\section{A \boldmath{$\simmass \gev$} Higgs boson in the S2HDM}

In this section we briefly summarize the
scalar sector of
S2HDM and how the excesses at about $\simmass \gev$
can be accommodated in this model.
We also review the relevant experimental
and theoretical constraints that are
applied in our numerical analysis.

\subsection{Model definitions}
\label{sec:modeldef}

The S2HDM extends the SM,
containing only one SU(2) Higgs doublet
$\Phi_1$, by a second
Higgs doublet field $\Phi_2$
and an additional
complex gauge-singlet field
$\Phi_S$~\cite{Jiang:2019soj,
Biekotter:2021ovi}.
As in the 2HDM, in the S2HDM the electroweak symmery is 
spontaneously broken by the two vacuum expectation values (vevs)
of the two Higgs doubles, $v_1$ and $v_2$, with
$\tan\beta = v_2 / v_1$,
and $v_1^2 + v_2^2 = v^2 \approx (246 \gev)^2$
corresponds to the SM vev squared.
In addition, the real component of the
singlet field has the non-zero vev $v_S$.
\TBB{Imposing a
softly-broken U(1) symmetry 
under which
only $\Phi_S$ is charged,
\hto{the imaginary part of the singlet} gives rise to the presence
of a stable pseudo-Nambu-Goldstone 
DM state\hto{. This yields} the attractive
possibility of accommodating the
observed DM relic abundance 
\GWnn{via} the usual
freeze-out mechanism.}
Neglecting possible sources of CP violation,
as we do throughout this paper for simplicity,
the physical scalar spectrum of the S2HDM consists
of three CP-even Higgs bosons $h_{1,2,3}$
with masses $m_{h_{1,2,3}}$,
a pair of charged Higgs bosons $H^\pm$ and
a CP-odd Higgs boson $A$ with masses
$m_{H^\pm}$ and $m_A$, respectively, as well as
a stable scalar~DM candidate~$\chi$.

We impose an additional 
$Z_2$ symmetry under which
one of the doublet fields changes the sign
in order to avoid flavor changing neutral
currents at the tree-level,
The $Z_2$~symmetry is only softly-broken via a term of
the form $-m_{12}^2(\Phi_1^\dagger \Phi_2
+ \mathrm{h.c.})$.
This symmetry 
implies for the fermion sector 
that either $\Phi_1$ or $\Phi_2$
(but not both) couples
to either the charged leptons $\ell$, the up-type
quarks $u$ or the down-type quarks $d$, leading to the
four Yukawa types of the model~\cite{Branco:2011iw}:
type~I, II, III (lepton-specific) and
IV (flipped).

\subsection{Interpretation of the excesses}
\label{sec:quanti}

In the following discussion,
the lightest of the three CP-even Higgs
bosons of the S2HDM $h_1$
serves as the possible particle state
at \GW{about} $\simmass \gev$, also denoted \hnf\ from here on.
We furthermore assume that
the second lightest Higgs boson, $h_2 = \hotf$,
corresponds to the state discovered
at about $125 \gev$.
The key aspect of the signal interpretation
presented here is that \hnf\ obtains
its couplings to the
fermions and gauge
bosons as a result of the mixing with the
CP-even components of the two doublets.
Despite the predominantly singlet-like character of \hnf, 
sizable decay rates into di-photon pairs can
be achieved via a suppression of the otherwise
dominating decay into $b$-quark 
pairs (see also \citere{Barbieri:2013nka}).
At the same time, 
no such suppression should occur for
the coupling to top quarks,
whose loop contribution gives rise to the
decay into photons
and also governs the production process via gluon fusion.
In the Yukawa types~II and~IV, $\Phi_1$
is coupled to down-type quarks
and $\Phi_2$ is coupled to up-type
quarks. In this case
an independent modification of the couplings
of the Higgs bosons $h_i$ to bottom quarks and
top quarks is possible.
These two types are therefore of particular interest 
regarding the prediction of a sufficiently large 
di-photon signal rate
(see \citere{Biekotter:2023jld} for a
detailed discussion).
\htr{On the other hand, between these two types, an important difference
arises from the fact that 
$c_{h_{95} \tau^+ \tau^-}$ is equal to
$c_{h_{95} b \bar b}$ in type~II, 
\GWn{while it is equal} 
to $c_{h_{95} t \bar t}$ in type~IV.
Accordingly, 
\TBn{in type~II no}
sizable signal rates in the $\tau^+ \tau^-$
decay channel \TBn{can be achieved}
if the di-photon excess
is accommodated, whereas in type~IV a larger rate
in the $\tau^+\tau^-$ 
\GWn{channel can occur simultaneously with a 
relatively large rate in the 
$\gamma\gamma$ channel}~\cite{Biekotter:2022jyr}.}

\subsection{Constraints}
\label{sec:constraints}

The parameter space that is relevant for
a possible description of the excesses at about
$\simmass \gev$ is subject to various theoretical
and experimental constraints.

Theoretical constraints that we apply in our analysis
ensure that the perturbative treatment of
the scalar sector of the S2HDM is valid
by demanding agreement with perturbative
unitarity constraints~\cite{Biekotter:2021ovi}.
In addition,
according to the approach
described in \citere{Biekotter:2021ovi}
we require that the tree-level scalar potential
is bounded from below, and that the
electroweak vacuum corresponds to the
global minimum of the
potential.

With regard to the 
experimental constraints,
we check whether the parameter points
are in agreement with the cross section
limits from BSM Higgs searches 
by making use of the public code
\texttt{HiggsBounds}~\cite{Bechtle:2008jh,
Bechtle:2011sb,
Bechtle:2013wla,
Bechtle:2020pkv,
Bahl:2022igd} (as part of the new code 
\texttt{HiggsTools v.1}~\cite{Bahl:2022igd}).
A parameter point is rejected if
the signal rate of one of the Higgs bosons
in the most sensitive search channel
(based on the expected limits) is larger
than the experimentally observed
limit at the 95\% confidence level.
In order to ensure that 
the properties of \hotf~are in agreement
with the measured signal rates from the LHC,
we make use of the public code
\texttt{HiggsSignals}~\cite{Bechtle:2013xfa,
Bechtle:2014ewa,
Bechtle:2020uwn,
Bahl:2022igd} (also part of 
\texttt{HiggsTools}~\cite{Bahl:2022igd}).
We regard a parameter point as accepted for 
$\chi^2_{125} \leq \chi^2_{{\rm SM},125} + 6.18$,
where $\chi^2_{125}$ is the value of
a $\chi^2$-fit to the various LHC cross-section
measurements in the S2HDM, and $\chi^2_{{\rm SM},125}$
is the SM fit result.
In two-dimensional parameter
\TB{estimations, in which other free parameter
are profiled over,}
the above condition ensures that the 
accepted S2HDM parameter points are not disfavoured
by more than $2\,\sigma$ in comparison to the SM
from the LHC rate measurements.

Flavor-physics
observables and from electroweak precision
observables give rise to indirect experimental
constraints on the Higg sector of the~S2HDM.
We apply lower limits of $\tan\beta > 1.5$ and
$m_{H^\pm} > 600\gev$
in our S2HDM parameter scans in type~II and type~IV
to ensure agreement with
the flavor-physics constraints~\cite{Haller:2018nnx}.
Regarding electroweak precision observabes,
we apply constraints in terms of the oblique parameters
$S$, $T$ and $U$, computed according to
\citere{Grimus:2008nb} at the one-loop level.
We impose that the predicted values of the
oblique parameters are in agreement with the 
fit result of \citere{Haller:2018nnx}
within~$2\,\sigma$ confidence
level.\footnote{We do not consider the recent
CDF measurement of $M_W$~\cite{CDF:2022hxs} here,
see the discussion in \citere{Biekotter:2022abc}.}

\TBB{Regarding the description of the
excesses at \hto{about} 95~GeV, 2HDM + singlet
models like the S2HDM provide a
rather generic framework representative
of a wide class of models with extended
scalar sectors. 
In our numerical analysis,}
we applied the Planck measurement of today's
relic abundance
of $h^2 \Omega = 0.119$~\cite{Planck:2018vyg}
\TBB{as an upper limit, hence preventing 
an overclosure of
the universe but
allowing for an
under-abundance of the predicted DM.
We used} \texttt{micrOMEGAs}~\cite{Belanger:2018ccd}
for the calculation of the predicted
relic abundance, \TBB{assuming a standard cosmological
history.
We additionally} applied \TBB{the most stringent}
DM direct-detection limits
\TBB{from the LUX-ZEPLIN experiment~\cite{LZ:2022lsv}
using the predictions for the DM-nucleon
scattering cross sections at one-loop level
computed in}
\citere{Biekotter:2022bxp}.
\TBB{We note that fulfilling the condition
of being in agreement with the applied DM constraints
and the condition of describing the di-photon
excess at~95~GeV can be satisfied 
\GWnn{simultaneously since the constraints mainly affect 
different \hto{sectors} of the model.}
The only exception is that the combined effect
of the constraint from the DM relic abundance,
the signal-rate measurements of the detected
Higgs boson at~125~GeV and demanding 
\GWnn{compatibility with}
the 95~GeV excesses excludes parameter points
for which the decay $h_{125} \to \chi \chi$ is
kinematically open.}

\section{Numerical analysis}
\label{sec:num}

In order to investigate
the impact 
of the increased sensitivity in
the di-photon channel due to the combination of
ATLAS and CMS data 
in the S2HDM interpretation, 
we performed a
parameter scan in the Yukawa types~II
and~IV of the S2HDM.
We analyzed
the theoretical predictions in comparison to 
the experimental results for the observed excesses near $95\gev$, 
ensuring at the same time that the properties of \hotf\
are in good agreement with the most
up-to-date LHC signal rate measurements.
We quantify the compatibility with
the excesses at $\massAC \gev$ using the 
contributions $\chi^2_{\gamma\gamma}$,
$\chi^2_{bb}$, and $\chi^2_{\tau\tau}$
which are defined as
\begin{equation}
\chi^2_{\gamma\gamma,\tau\tau,bb} =
\frac{
(\mu_{\gamma\gamma,\tau\tau,bb} -
\mu_{\gamma\gamma,\tau\tau,bb}^{\rm exp})^2 }{
(\Delta \mu_{\gamma\gamma,\tau\tau,bb}^{\rm exp})^2} \ .
\label{eq:chisq95indi}
\end{equation}
For the experimental central
values and the uncertainties 
we use the values
stated in \refse{sec:intro}, and
$\mu_{\gamma\gamma,\tau\tau,bb}$
are the theoretically predicted values.
Since $\mu_{\gamma\gamma}^{\rm exp}$ has
asymmetric uncertainties, we define
$\chi^2_{\gamma\gamma}$ in such a way that
the lower uncertainty is used if
$\mu_{\gamma\gamma} < \mu_{\gamma\gamma}^{\rm exp}$,
and the upper uncertainty is used if
$\mu_{\gamma\gamma} > \mu_{\gamma\gamma}^{\rm exp}$.
To obtain the predictions for
$\mu_{\gamma\gamma}$ and
$\mu_{\tau\tau}$,
we used \texttt{HiggsTools}
to derive the gluon-fusion cross section
of the state at $95\gev$
via a re-scaling of the SM predictions
as a function of $c_{h_{95} t \bar t}$ and
$c_{h_{95} b \bar b}$. To compute
$\mu_{b b}$, we approximated the cross section
ratio as
$\sigma / \sigma_{\rm SM} = c_{h_{95} VV}^2$.
The branching ratios
of \hnf\ were obtained with the help
of \texttt{N2HDECAY}~\cite{Muhlleitner:2016mzt,
Engeln:2018mbg}.
For the generation of the S2HDM parameter
points and the application of the
constraints, we used the program
\texttt{s2hdmTools}~\cite{Biekotter:2021ovi,
Biekotter:2022bxp}, which features interfaces
to \texttt{HiggsTools}, 
\texttt{micrOMEGAs} and \texttt{N2HDECAY}.

\TBB{

\subsection{Genetic algorithm}

We scanned the S2HDM parameter space using
a genetic algorithm in order to 
\GWnn{determine the parameter regions that are}
suitable
for the description of the di\hto{-}photon excess while
being in agreement with the various theoretical and
experimental constraints \hto{discussed above.}
Genetic algorithms mimic the
process of natural selection to find solutions
\GWnn{for} problems whose solution space can be quantified
in terms of a fitness function 
\GWnn{that needs}
to be maximized.
The application
of the genetic algorithm significantly improved the
running time required to find a valid parameter point
compared to a random sampling of the model parameters.
Compared to other optimization techniques,
e.g.~Markov Chain Monte Carlo algorithms
or multimodal nested sampling algorithms,
genetic algorithms are
particularly well suited for the task at \GWnn{hand.
This is due to the fact that}
they are derivative-free and can be applied to
non-differentiable optimization problems as present here due
to discrete constraints, such as the \GWnn{boundedness}-from-below
conditions or the 95\% confidence-level cross section
limits from BSM scalar searches.
In addition, genetic algorithms are easily
parallelizable,
and they are particularly well suited to uncover
novel and unconventional solutions to a problem as a result
of the mating and mutation steps,
as \hto{detailed} below.
Before \GWnn{discussing} the specifics of the algorithm used
in our analysis we state the
scan ranges of the free parameters.

The mass of \hnf\
\GWnn{has been} varied in the region in which the
di-photon excess is most pronounced,
i.e.~$94\gev \leq m_{\hnf} \leq 97\gev$. 
The mass of the second-lightest Higgs boson
\GWnn{has been} set to $m_{\hotf} = 125.09\gev$.
The masses of the remaining
neutral and charged
Higgs bosons, denoted \GWnn{as} $H$, $A$ and
$H^\pm$
in the following,
as well as the mass of
the DM~state, 
\GWnn{have been} scanned up to an
upper limit of \hto{$1 \tev$.}
For the mass of~$H^\pm$ additionally
the lower limit $m_{H^\pm} > 600\gev$ 
\GWnn{has been}
applied
due to flavor-physics constraints,
see above.
Moreover, we \GWnn{have} varied $\tan\beta$ in the
range $1.5 \leq \tan\beta\leq 10$, and for the
singlet vev we \GWnn{have chosen} $40\gev \leq v_S \leq 2\tev$.
Finally, the scan range of the parameter
$m_{12}^2$ \GWnn{has been} determined by the condition
$400\gev \leq M \leq 1\tev$, where
$M^2 = m_{12}^2 / (\sin\beta \cos\beta)$.

For the implementation of the genetic algorithm
we used the software package
\texttt{DEAP}~\cite{DEAP_JMLR2012}.
\GWnn{As starting point for obtaining parameter points}
that describe the
observed excesses and which are in agreement with
the constraints, we created a so-called
population of 500,000
parameter points, called individuals, where each
individual is defined by a list of floating point
numbers corresponding to the values of the
free parameters. This list can be thought of as the
chromosome of an individual.
The chromosomes of the individuals of the initial
population were generated by \GWnn{randomly} assigning 
the values of the free parameters within the
scan ranges given above.

The goodness-of-fit of each
parameter point is quantified in terms of a
so-called fitness function whose value becomes larger
with increasing quality of the fit.
We defined a fitness function depending on the
compatibility with the observed excesses in terms
of $\chi^2_{\gamma\gamma}$, $\chi^2_{\tau\tau}$
and $\chi^2_{bb}$, on the compatibility with the
signal-rate measurements of the Higgs boson
at~125~GeV in terms of $\chi^2_{125}$,
and by piece-wise assigning large penalties to the
fitness if any of the theoretical or \hto{other} experimental
constraints were violated.
The genetic algorithm
operates by progressively generating new
generations of individuals through combining
the traits of individuals from the previous generation.
This \textit{crossover} procedure is
defined in such a way
\GWnn{that}
the fitness of the individuals from
one generation to the other \GWnn{is improved}, and
it works in three steps:

The first step is called
selection, where based on the fitness of the individuals
a subset of the population is selected that is
allowed to participate in the production of new
individuals, called off-springs. In our parameter scan,
we used tournament selection with size $k=3$,
i.e.~from the existing population three individuals
are randomly drawn and only the individual with the
best fitness is selected to be able to participate
in the subsequent step.
\GWnn{This is done until a selected population of 500,000 is achieved.}
Since the same individual can be selected more than
once, the number of selected individuals
is in general smaller than the total number
of individuals in the \GWnn{original} population.

The second step of the crossover
procedure is the mating stage.
The mating stage creates new individuals
whose chromosomes are inherited from
the subset
of the previously selected individuals. We used
uniform crossover with two parent individuals, where
each element of the chromosome of the off-spring
is taken over randomly from one of the two parent
individuals. By empirically studying the rate
of increase of the fitness values, we found that
the performance of the algorithm is improved if
a dominant part of the parameter values is taken over
from one of the \GWnn{parents}.
Conversely, 
mixing the parameter values
from both parents in equal amounts often 
results in 
off-springs with poor 
fitness, where the corresponding parameter
point violates at least one of the theoretical or
experimental constraints due to significant changes
in too many parameters.
We therefore set a probability of~80\% for each
element of the chromosome of the off-spring
to come from one of the parents, and 20\%
from the other parent.
With the mating procedure we produced a total
number of 500,000 off-springs.

The third and final step of the crossover procedure is
the mutation stage. At this state, a 
randomly chosen subset
of the new population is \GWnn{exposed} to 
a
modification of their chromosomes. The mutation
stage is especially important to maintain diversity
in the population, and thus for the coverage of
a large region of the solution space by preventing
that the algorithm becomes trapped in 
local maxima of the fitness function.
In our scan, 10\% of the population was chosen
to be \GWnn{exposed} to mutation. 
The chromosome of an
individual was mutated by replacing 
\hto{each} element
with a probability of~10\% with a float number
that was randomly chosen within an interval
between 0.8 and 1.2 times the original value.
After the mutation stage, the production of the
new generation of the population is complete.

We let the algorithm evolve for \GWnn{a} total number
of 40 generations. In order \GWnn{not to} lose the
individual with the best fitness at intermediate
stages due to the mating or mutation steps,
the best-fit individual was always appended
separately to the new generation of individuals
without any modification to its chromosome.
At the end of the evolution, the individual
with the best fitness was saved, i.e.~we ran
the genetic algorithm once for each
parameter point contained in our final sample \GWnn{displayed in the plots.}
Each \GWnn{run} of the genetic algorithm 
took about one hour on a typical CPU.
Running the algorithm in parallel on
a 48-CPU cluster enabled us to
efficiently obtain the final sample of
parameter points discussed in our
analysis within a few days.
These points adhere to all applied
constraints and have been selected
based on their compatibility with the
observed excesses at 95~GeV.
}

\subsection{Analysis of the di-photon excess}

\begin{figure*}[t]
\centering
\includegraphics[width=0.9\columnwidth]{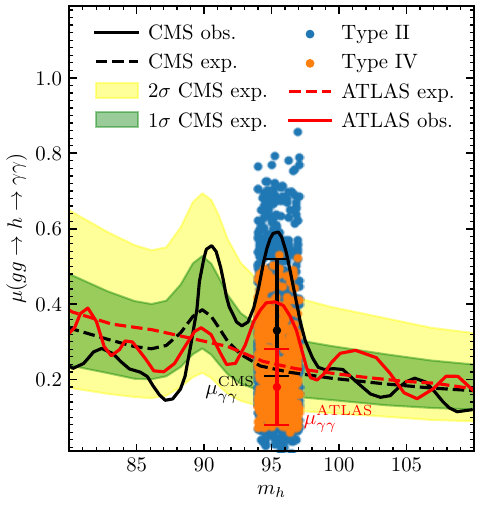}~
\includegraphics[width=0.9\columnwidth]{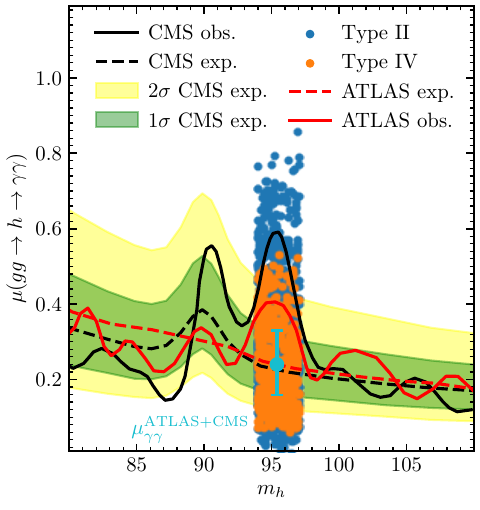}~
\vspace*{-0.4cm}
\caption{\small
S2HDM parameter points
passing the applied constraints
in the \plane{m_{h_{1}}}{\mu_{\gamma\gamma}}
for the type~II (blue) and the type~IV (orange).
The expected and observed
cross section limits obtained
by CMS are indicated by
the black dashed and solid lines, respectively,
and the $1\sigma$ and $2\sigma$ uncertainty intervals
are indicated by
the green and yellow
bands, respectively.
Overlaid in red are the expected and observed 
\GW{limits from}
ATLAS~\cite{ATLAS-CONF-2023-035}. 
The values of $\muAgaga$, $\muCgaga$ and 
$\muACgaga$ and their respective uncertainties
are indicated by
the red, \TB{black} \SH{(left plot)} 
and cyan \SH{(right plot)} error bars at 
\GW{$\massAC \gev$}.}
\label{fig:gamgam}
\end{figure*}

In \reffi{fig:gamgam} we show the
predictions for $\mu_{\gamma\gamma}$ for the
S2HDM parameter points that are in agreement
with the applied constraints.
The type~II parameter points are shown
in blue, and the parameter points of type~IV
are shown in orange. 
\GW{It should be noted that the orange points are plotted above 
the blue ones, i.e.\ the whole range displayed for the orange points 
is also covered by the blue points.}
The expected and observed
cross section limits 
obtained by CMS are indicated by
the black dashed and solid lines, respectively.
The $1\sigma$ and $2\sigma$ uncertainty bands
are indicated by the green and yellow
bands, respectively~\cite{CMSnew}.
Overlaid are the expected and observed 
95\% confidence-level limits
\TBn{on the signal strengths
observed} by ATLAS~\cite{ATLAS-CONF-2023-035} 
as dashed and solid red lines, respectively.
\TBn{We obtained these limits by normalizing the
expected and observed
cross-section limits reported by ATLAS
with the cross sections predicted for
a SM Higgs boson at the same
mass~\cite{LHCHiggsCrossSectionWorkingGroup:2016ypw}
using \texttt{HiggsTools}~\cite{Bahl:2022igd}.}
The values of $\muAgaga$, $\muCgaga$ and 
$\muACgaga$ and their respective uncertainties
are indicated by
the red, black \SH{(left plot)} and 
cyan \SH{(right plot)} error bars at 
\GW{$\massAC \gev$}.
One can see that both types of the S2HDM
considered here \GW{can accommodate the combined observed
excess}. 
Type~II can give rise to larger predicted
values of $\mu_{\gamma\gamma}$ due to a 
suppression of the $h_{95} \to \tau^+ \tau^-$
decay mode,
see the discussion in \citere{Biekotter:2023jld}.

\subsection{Di-photon vs.\ \boldmath{$b \bar b$} vs.\ \boldmath{$\tau^+\tau^-$} excesses}
\label{sec:all3}

\begin{figure*}[t]
\centering
\includegraphics[width=0.82\columnwidth]{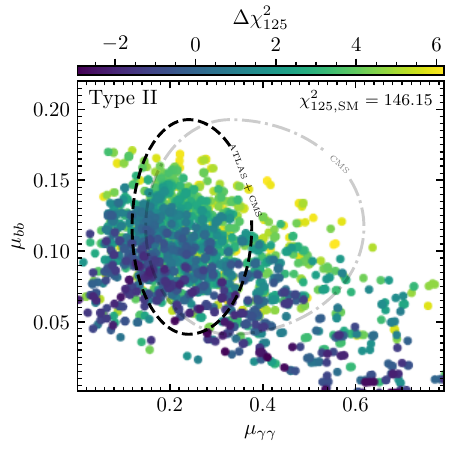}~~~
\includegraphics[width=0.82\columnwidth]{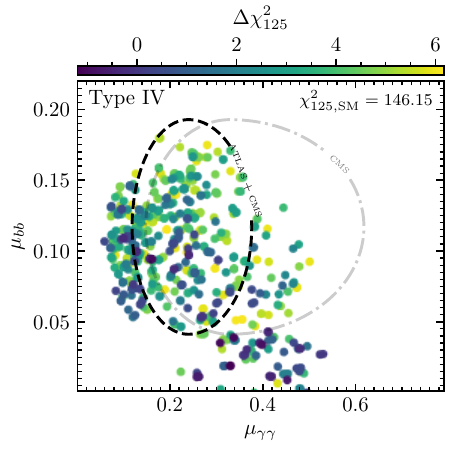}\\[0.4em]
~~\includegraphics[width=0.82\columnwidth]{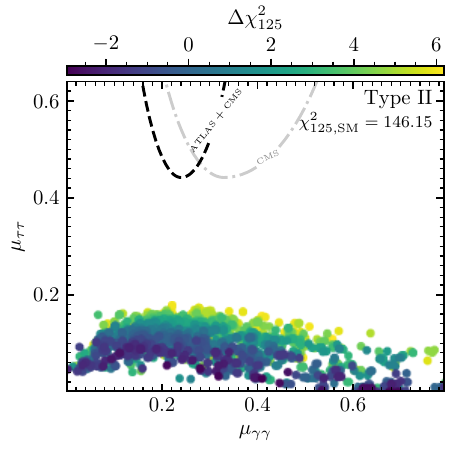}~~~
\includegraphics[width=0.82\columnwidth]{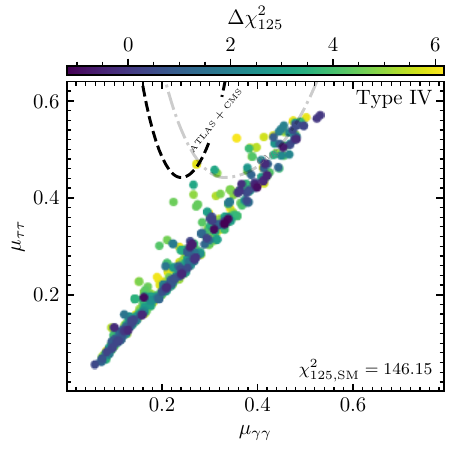}
\vspace*{-0.4cm}
\caption{\small
S2HDM parameter points 
passing the applied constraints
in the \plane{\mu_{\gamma\gamma}}{\mu_{bb}}
(top row) and the
\plane{\mu_{\gamma\gamma}}{\mu_{\tau\tau}}
(bottom row) for type~II (left)
and type~IV (right). The colors of the
points indicate the value of $\Delta \chi^2_{125}$.
The black dashed lines indicate the
regions in which the two excesses considered
in each plot are accommodated at a level of
$1\sigma$ or better, i.e.~$\chi^2_{\gamma\gamma}
+ \chi^2_{bb} \leq 2.3$ (top row) and
$\chi^2_{\gamma\gamma}
+ \chi^2_{\tau\tau} \leq 2.3$ (bottom row).
\SH{The corresponding gray dot-dashed lines indicate
the previous result based solely on the CMS Run~2 data.}}
\label{fig:yyllbb}
\end{figure*}

In the previous subsection
we demonstrated that
both the Yukawa types~II and~IV
can describe the
excess in the di-photon channel observed
by ATLAS and CMS. Now we turn to the question whether
additionally also the $b \bar b$ excess
observed at LEP and/or the $\tau^+ \tau^-$
excess at CMS can be accommodated.

Starting with the $b \bar b$ excess,
we show in the top row of \reffi{fig:yyllbb}
the parameter points 
passing the applied constraints in the 
\plane{\mu_{\gamma\gamma}}{\mu_{b b}}.
The parameter points of type~II and type~IV
are shown in the left and the right plot, respectively.
The colors of the points indicate the
value of $\Delta \chi^2_{125}$, quantifying
the degree of compatibility with the LHC rate
measurements of $h_{125}$.
The black dashed lines indicate the
region in which the excesses are described
at a level of $1\sigma$ or better,
i.e.~$\chi^2_{\gamma\gamma} +
\chi^2_{bb} \leq 2.3$ 
(see \refeq{eq:chisq95indi}).
\SH{The corresponding gray dot-dashed lines indicate the 
previous result based solely on the CMS Run~2 data
regarding the di-photon excess.}

One can observe that there are points
inside the $1\,\sigma$ preferred region 
in the upper left and right plots.
Thus, both type~II and type~IV
are able to describe the increased sensitivity in
the di-photon channel, now reaching $\sigAC\,\sigma$, 
and the $b \bar b$ excess
simultaneously.
At the same time the properties
of the second-lightest scalar
\hotf\ are such that the
LHC rate measurements
can be accommodated 
at the same $\chi^2$ level as in the SM,
i.e.~$\Delta \chi^2_{125} \approx 0$,
or 
better. Such points are found 
inside the $1\,\sigma$ preferred region 
\GW{for $\mu_{bb}$ values below the central value}. 
At the current level of experimental
precision, the description of both
excesses is therefore possible in combination
with the presence of
a Higgs boson at $125\gev$ that
would so far be indistinguishable from
a SM Higgs boson.

Turning to the di-tau excess,
we show in the bottom row of
\reffi{fig:yyllbb} the 
parameter points 
passing the applied constraints
in the \plane{\mu_{\gamma\gamma}}{\mu_{\tau\tau}}.
As before, the colors of the points
indicate the values of $\Delta \chi^2_{125}$.
The black dashed lines 
indicate the region in which the di-photon excess and
the di-tau excess are described at a level
of $1\,\sigma$ or better,
i.e.~$\chi^2_{\gamma\gamma} + \chi^2_{\tau\tau} \leq 2.3$,
whereas the gray dot-dashed line corresponds
to the preferred $1\sigma$ region based solely on
the CMS result regarding the di-photon excess.

In the lower left plot, showing
the parameter points of the scan in
type~II, one can see that there are no
points within or close to the $1\,\sigma$ region.
This finding 
\GWn{can be understood from the discussion in \refse{sec:quanti}. The conclusion that the two excesses cannot be simultaneously described in this case}
is qualitatively unchanged as compared to
the results of \citeres{Biekotter:2022jyr,Biekotter:2023jld},
where $\mu_{\gamma\gamma}^{\rm exp} = 0.6 \pm 0.2$ and
$0.33^{+0.19}_{-0.12}$ were used, respectively.

The lower right plot shows the 
parameter points 
passing the applied constraints
from the
scan in type~IV. One can observe that
the values of $\mu_{\tau\tau}$ overall increase
with increasing \GWn{values} of $\mu_{\gamma\gamma}$.
\GWn{While some parameter points}
reach the lower edge of the black $1\,\sigma$ line,
\GWn{the new result for $\mu_{\gamma\gamma}$ worsens 
the simultaneous compatibility with the di-photon 
and the di-tau excess for the type~IV case as 
compared to the previous result for the di-photon excess
that was based solely on the CMS result. Regarding the 
di-tau excess,}
all points lie substantially
below the central value of $\mu_{\tau\tau}^{\rm exp}$.
Although larger values of $\mu_{\tau\tau}$
can theoretically be achieved in
type~IV~\cite{Biekotter:2022jyr},
such parameter points are
excluded by experimental bounds from
recent searches performed by CMS 
for the production of a Higgs boson
in association with a top-quark pair or
in association with a $Z$~boson, with subsequent
decay into tau pairs~\cite{CMS-PAS-EXO-21-018}.
\TBB{In addition,
in the S2HDM values of $\mu_{\tau\tau} \gtrsim 0.7$
are in tension with cross section limits
from Higgs-boson searches at LEP 
\GWnn{for the decay of the Higgs boson} into a pair of
$\tau$-leptons~\cite{Barate:2003sz,
Biekotter:2022jyr}.}
Consequently, a simultaneous description
of 
the $\gamma\gamma$ and the $\tau\tau$ 
excesses is 
possible 
\GWn{at best} at the level of $1\,\sigma$.
We note here that a better description of both
the di-photon and the di-tau excess can
be achieved 
\GWn{if $h_{95}$ is identified with a 
CP-odd state~\cite{Azevedo:2023zkg},
because such a scenario is less}
constrained by the limits 
\GWn{arising from}
top-quark associated
production
(see also \citere{Iguro:2022dok}).

\TBB{
In \refta{tab:bp2} we provide details
of a selection of three benchmark points 
that we obtained in our parameter scan
in the type~II S2HDM.
These benchmark points 
\GWnn{feature} 
a very good description of the di-photon excess
observed at the LHC in combination with
the $b \bar b$ excess observed at LEP, while the
excess of di-tau events observed by CMS
cannot be described in type~II 
\GWnn{as discussed}
above.
Moreover, the benchmark points BP1 and BP3 saturate
the measured DM relic abundance,
while the DM density predicted for BP2
is under-abundant, leaving room for additional
components contributing
to the observed DM relic abundance.
For BP1 the
DM state $\chi$ has a mass of $m_\chi = 63.3\gev$,
thus annihilating efficiently via
$s$-channel exchange of $h_{125}$, while
the invisible decay $h_{125} \to \chi \chi$
is still kinematically
closed \GWnn{for the on-shell case}.\footnote{\TBB{Such 
a scenario is
especially compelling in
view of the
excess of gamma rays from the
galactic center observed by the Fermi Large Area
Telescope~\cite{Fermi-LAT:2010cni,
Fermi-LAT:2016uux,Fermi-LAT:2017opo}.
\GWnn{A possible} origin of this long-standing excess
\GWnn{could}
be the annihilation
of a
WIMP DM candidate in this mass
window~\cite{Hooper:2010mq,Daylan:2014rsa},
potentially in agreement with the
properties of the DM state~$\chi$ for BP1 (see \citere{Biekotter:2021ovi} for a detailed
discussion).}}
For BP3, $\chi$ is substantially heavier,
$m_\chi = 964\gev$, and annihilates mainly via
processes involving the heavy BSM states
$h_3$ \hto{and} $A$.
A typical feature of the parameter points
describing the di-photon excess is the suppression
of the couplings of $h_{95}$ and $h_{125}$
to down-type quarks compared to the couplings
to up-type quarks. Regarding the phenomenology
of the heavier scalars, typically the third
CP-even scalar $h_3$ is lighter than
the 
\GWnn{CP-odd state} $A$ and the charged scalars
$H^\pm$. If the mass splitting is sufficiently
large, e.g.~for BP2,
\GWnn{searches for} signals like $A \to Zh_3$ and
$H^\pm \to W^\pm h_3$ can be utilized to probe
the preferred parameter space regions.
\GWnn{This is of particular interest since}
the parameter space regions 
\GWnn{where these channels are kinematically open are also}
favoured
by the presence of a first-order EW phase transition
in SM extensions containing a second
Higgs doublet~\cite{Biekotter:2023eil}.
Due to the required departures from the alignment
limit in order to produce sizable signal rates
for the state $h_{95}$, 
the parameter space relevant for
a description of the di-photon excess
\GWnn{may also be probed}
via searches for otherwise strongly
suppressed decays of~$A \to Z h_{125}$,
or searches for the decay $H \to h_{125} h_{125}$.
For the latter,
we find branching ratios at the level
of 
\GWnn{10\%}, see BP1 and BP2
in \refta{tab:bp2}.
}

\begin{table*} 
{\footnotesize
\renewcommand*{\arraystretch}{1.2}
{\bf \ Benchmark points: Type~II} \\[0.4em]
\begin{tabular}{l|cccccccccccc}
\textbf{Parameters} &
  $\tan\beta$ & $\alpha_1$ & $\alpha_2$ & $\alpha_4$ &
  $m_{h_1}$ & $m_{h_2}$ & $m_{h_3}$ & $m_A$ &
  $m_{H^\pm}$ & $m_\chi$ & $v_s$ & $M$ \\
\hline
\makebox[\widthof{\textbf{Effective couplings}}][l]{BP1}
  & 2.79 & 1.35 & 1.22 & 1.49 &
  95.7 & 125 & 687 & 812 & 658 & 63.3 & 1674 & 664 \\
BP2 & 3.65 & 1.39 & 1.20 & 1.52 & 95.1 &
  125 & 556 & 681 & 669 & 276 & 843 & 548 \\
BP3 & 3.30 & 1.41 & 1.25 & 1.46 & 95.9 & 125 &
  785 & 848 & 849 & 964 & 600 & 770
\end{tabular}\\[0.4em]
\begin{tabular}{l|ccccc}
\textbf{Phenomenology} &
  $\mu_{\gamma\gamma}$ & $\mu_{bb}$ &
  $\mu_{\tau\tau}$ & $\Delta \chi^2_{125}$ &
  $h^2 \Omega$ \\
\hline
\makebox[\widthof{\textbf{Effective couplings}}][l]{BP1}
  & 0.247 & $9.94 \cdot 10^{-2}$ &
  $9.62 \cdot 10^{-2}$ & -0.467 & 0.114 \\
BP2 & 0.245 & 0.118 & 0.109 & 2.10 & $4.67 \cdot 10^{-3}$ \\
BP3 & 0.243 & $7.73 \cdot 10^{-2}$ & $7.43 \cdot 10^{-2}$ &
  1.74 & 0.104
\end{tabular}\\[0.4em]
\begin{tabular}{l|cccccc}
\textbf{Branching ratios} \\
$\bm{h_1}$ & $b \bar b$ & $\tau^+ \tau^-$ & $WW$ & $ZZ$ &
  $\gamma\gamma$ \\
\hline
\makebox[\widthof{\textbf{Effective couplings}}][l]{BP1}
  & 0.690 & $6.99\cdot 10^{-2}$ & $1.01\cdot 10^{-2}$ &
    $1.38\cdot 10^{-3}$ & $2.74\cdot 10^{-3}$ \\
BP2 & 0.717 & $7.27 \cdot 10^{-2}$ & $8.39 \cdot 10^{-3}$ &
  $1.20 \cdot 10^{-3}$ & $2.44 \cdot 10^{-3}$ \\
BP3 & 0.644 & $6.52 \cdot 10^{-2}$ & $1.26 \cdot 10^{-2}$ &
  $1.70 \cdot 10^{-3}$ & $3.26 \cdot 10^{-3}$ \\
\hline
\hline
$\bm{h_2}$ & $b \bar b$ & $\tau^+ \tau^-$ & $WW$ & $ZZ$ &
  $\gamma\gamma$ \\
\hline
BP1 & 0.543 & $5.83 \cdot 10^{-2}$ & 0.238 & $2.98 \cdot 10^{-2}$ &
  $2.61 \cdot 10^{-3}$ \\
BP2 & 0.532 & $5.71 \cdot 10^{-2}$ & 0.246
  & $3.08 \cdot 10^{-2}$ & $2.61 \cdot 10^{-3}$ \\
BP3 & 0.570 & $6.52 \cdot 10^{-2}$ &
  0.222 & $2.78 \cdot 10^{-2}$ & $2.41 \cdot 10^{-3}$ \\
\hline
\hline
$\bm{h_3}$ & $t \bar t$ & $h_2 h_2$ & $h_1 h_2$ &
  $WW$ & $ZZ$ & $h_1 h_1$ \\
\hline
BP1 & 0.688 & 0.117 & $7.09 \cdot 10^{-2}$ &
  $5.93 \cdot 10^{-2}$ & $2.89 \cdot 10^{-2}$ &
  $9.32 \cdot 10^{-3}$ \\
BP2 & 0.631 & 0.123 & $6.24 \cdot 10^{-2}$ &
  $5.71 \cdot 10^{-2}$ & $2.73 \cdot 10^{-2}$ &
  $1.06 \cdot 10^{-2}$ \\
BP3 & 0.766 & $5.27 \cdot 10^{-2}$ &
  $7.06 \cdot 10^{-2}$ & $2.99 \cdot 10^{-2}$ &
  $1.47 \cdot 10^{-2}$ & $2.05 \cdot 10^{-2}$ \\
\hline
\hline
$\bm{A}$ & $W^\pm H^\mp$ & $t \bar t$ & $Z h_3$ &
  $Z h_1$ & $Z h_2$ \\
\hline
BP1 & 0.566 & 0.313 & $8.25 \cdot 10^{-2}$ &
  $1.88 \cdot 10^{-2}$ & $1.21 \cdot 10^{-2}$ \\
BP2 &  & 0.604 & 0.312 & $2.21 \cdot 10^{-2}$ &
  $2.33 \cdot 10^{-2}$ \\
BP3 &  & 0.881 & $7.43 \cdot 10^{-2}$ &
  $7.92 \cdot 10^{-3}$ & $3.33 \cdot 10^{-3}$ \\
\hline
\hline
$\bm{H^\pm}$ & $tb$ & $W h_1$ & $W h_2$ & $W h_3$ \\
\hline
BP1  & 0.936 & $3.68 \cdot 10^{-2}$ & $2.32 \cdot 10^{-2}$ \\
BP2 & 0.657 & $2.33 \cdot 10^{-2}$ &
  $2.45 \cdot 10^{-2}$ & 0.289 \\
BP3 & 0.908 & $7.68 \cdot 10^{-2}$ &
  $8.28 \cdot 10^{-3}$
\end{tabular}\\[0.4em]
\begin{tabular}{l|ccc|ccc|ccc}
\textbf{Effective couplings} & $c_{h_1 VV} $ &
  $c_{h_1 t \bar t}$ & $c_{h_1 b \bar b}$
  & $c_{h_2 VV} $ &
  $c_{h_2 t \bar t}$ & $c_{h_2 b \bar b}$
  & $c_{h_3 VV} $ &
  $c_{h_3 t \bar t}$ & $c_{h_3 b \bar b}$\\
\hline
BP1 & 0.340 & 0.355 & 0.221 & -0.939 & -0.952 & -0.843 &
    $4.69 \cdot 10^{-2}$ & -0.311 & 2.83 \\
BP2 & 0.363 & 0.371 & 0.251 &
  -0.931 & -0.940 & -0.813 &
  $4.12 \cdot 10^{-2}$ & -0.232 & 3.689 \\
BP3 & 0.310 & 0.322 & 0.177 &
  -0.950 & -0.955 & -0.906 &
  $2.52 \cdot 10^{-2}$ & -0.277 & 3.324
\end{tabular}
\caption{\small Selection of
benchmark points from the scan in type~II
which describe the di-photon excess observed at
the LHC and the $b \bar b$ excess observed at LEP,
\TBB{while being in agreement with
all other experimental and theoretical constraints.}
BP1 and BP3 additionally
\TBB{saturate}
the measured
relic abundance of~DM, whereas for BP2 the
predicted DM density is under-abundant.}
\label{tab:bp2}
}
\end{table*}

\TBB{
In \refta{tab:bp4} we provide a similar selection
of benchmark points from the scan in the
S2HDM \GWnn{of} type~IV.
Compared to the type~II parameter points,
these parameter points feature also sizable
signal rates of~$h_{95}$ in the di-tau decay mode.
However, 
\GWnn{as discussed} above, the
predicted signal rates of $\mu_{\tau\tau}
\approx 0.4$ are still substantially
below the experimentally observed
value of $\mu_{\tau\tau}^{\rm exp} = 1.2 \pm 0.5$,
while at the same time the signal rates
for the di-photon excess $\mu_{\gamma\gamma}$
are predicted to
be slightly larger than the best-fit values 
\GWnn{in this parameter region}
(as is also visible
in the lower right plot of \reffi{fig:yyllbb}).
The parameter space regions preferred for a
description of the di-photon excess are
overall similar to the ones found for type~II,
with the exception that in type~IV it is possible
to reach larger values of $\tan\beta$.
This is
a consequence of the fact that in type~IV
the \GWnn{bounds from} 
LHC searches for heavy Higgs bosons decaying
into $\tau$-lepton pairs are
\GWnn{weaker}~\cite{ATLAS:2020zms,CMS:2022goy},
because the couplings
of~$A$ and of~$H$ to charged leptons
are suppressed
by factors of $1 / \tan\beta$, whereas
these couplings are enhanced in type~II
by factors of $\tan\beta$.
No relevant distinction 
\GWnn{occurs between the two}
types concerning the DM phenomenology.
As in type~II, the predicted DM relic density
can account for the observed \GWnn{DM} abundance 
\GWnn{(or of a large part of it)}, e.g.~for BP1 and BP3 shown in
\refta{tab:bp4}, while at the same time
the di-photon excess and the $b \bar b$ excess
observed at LEP can be described in
good agreement with the observed signal rates.
Still, it is also possible that the predicted
DM abundance is orders of magnitude below
the observed one, for instance for BP2,
where the DM mass $m_\chi = 276\gev$ is
very close to half the mass of the heavy
CP-even scalar, $m_{h_3} = 556\gev$, such that
the annihilation of~$\chi$ in the early universe
is resonantly enhanced 
\GWnn{(the branching ratio for
the decay $h_3 \to \chi\chi$ is also given in \refta{tab:bp4}).}
}

\begin{table*} 
{\footnotesize
\renewcommand*{\arraystretch}{1.2}
{\bf \ Benchmark point: Type~IV} \\[0.4em]
\begin{tabular}{l|cccccccccccc}
\textbf{Parameters} &
  $\tan\beta$ & $\alpha_1$ & $\alpha_2$ & $\alpha_4$ &
  $m_{h_1}$ & $m_{h_2}$ & $m_{h_3}$ & $m_A$ &
  $m_{H^\pm}$ & $m_\chi$ & $v_s$ & $M$ \\
\hline
\makebox[\widthof{\textbf{Effective couplings}}][l]{BP1}
  & 7.38 & -1.50 & -1.18 & -1.39 &
  95.2 & 125 & 918 & 903 & 834 & 517 & 674 & 918 \\
BP2 & 4.19 & 1.44 & 1.15 & 1.51 & 95.9 &
  125 & 742 & 785 & 778 & 314 & 132 & 739 \\
BP3 & 3.14 & 1.43 & 1.19 & 1.44 &
  95.2 & 125 & 578 & 838 & 842 & 312 &
  488 & 580
\end{tabular}\\[0.4em]
\begin{tabular}{l|ccccc}
\textbf{Phenomenology} &
  $\mu_{\gamma\gamma}$ & $\mu_{bb}$ &
  $\mu_{\tau\tau}$ & $\Delta \chi^2_{125}$ &
  $h^2 \Omega$ \\
\hline
\makebox[\widthof{\textbf{Effective couplings}}][l]{BP1}
  & 0.304 & $9.34 \cdot 10^{-2}$ & 0.353 & 0.227 & 0.0777 \\
BP2 & 0.382 & 0.117 & 0.412 & 3.03 & $1.22 \cdot 10^{-2}$ \\
BP3 & 0.343 & $7.50 \cdot 10^{-2}$ &
  0.414 & 5.55 & 0.107
\end{tabular}\\[0.4em]
\begin{tabular}{l|ccccccc}
\textbf{Branching ratios} \\
$\bm{h_1}$ & $b \bar b$ & $\tau^+ \tau^-$ & $WW$ & $ZZ$ &
  $\gamma\gamma$ \\
\hline
\makebox[\widthof{\textbf{Effective couplings}}][l]{BP1}
  & 0.544 & $0.183$ & $9.75\cdot 10^{-3}$ &
    $1.38\cdot 10^{-3}$ & $2.66\cdot 10^{-3}$ \\
BP2 & 0.554 & 0.189 & $1.15 \cdot 10^{-2}$ &
  $1.54 \cdot 10^{-2}$ & $3.01 \cdot 10^{-3}$ \\
BP3 & 0.458 & 0.231 & $1.18 \cdot 10^{-2}$ &
  $1.66 \cdot 10^{-3}$ & $3.21 \cdot 10^{-3}$ \\
\hline
\hline
$\bm{h_2}$ & $b \bar b$ & $\tau^+ \tau^-$ & $WW$ & $ZZ$ &
  $\gamma\gamma$ \\
\hline
BP1 & 0.514 & $7.54 \cdot 10^{-2}$ & 0.247 & $3.09 \cdot 10^{-2}$ &
  $2.76 \cdot 10^{-3}$\\
BP2 & 0.511 & $7.64 \cdot 10^{-2}$ & 0.247 &
  $3.09 \cdot 10^{-2}$ & $2.68 \cdot 10^{-2}$ \\
BP3 & 0.544 & $7.14 \cdot 10^{-2}$ & 0.231 &
  $2.89 \cdot 10^{-2}$ & $2.41 \cdot 10^{-2}$ \\
\hline
\hline
$\bm{h_3}$ & $t \bar t$ & $h_2 h_2$ & $h_1 h_2$ &
  $WW$ & $h_1 h_1$ & $W^\pm H^\mp$ & $\chi\chi$ \\
\hline
BP1 & $8.64 \cdot 10^{-2}$ & 0.181 & $0.299$ &
  $0.145$ & 
  $7.38 \cdot 10^{-4}$ & $2.24 \cdot 10^{-2}$ & \\
BP2 & 0.376 & 0.206 & $6.43 \cdot 10^{-2}$ &
  $9.56 \cdot 10^{-2}$ & 0.113 &  &
  $3.12 \cdot 10^{-2}$ \\
BP3 & 0.637 & 0.103 & 0.103 & $5.52 \cdot 10^{-2}$ &
  $3.45 \cdot 10^{-2}$ \\
\hline
\hline
$\bm{A}$ & $b \bar b$ & $t \bar t$ & $Z h_1$ &
  $Z h_2$ & $Z h_3$ \\
\hline
BP1 & 0.226 & 0.262 & $0.383$ &
  $2.39 \cdot 10^{-2}$ \\
BP2 & $7.03 \cdot 10^{-2}$ & 0.773 &
  0.102 & $5.24 \cdot 10^{-2}$ \\
BP3 & $4.08 \cdot 10^{-3}$ & 0.143 &
  $2.42 \cdot 10^{-2}$ & $4.23 \cdot 10^{-2}$ &
  0.824 \\
\hline
\hline
$\bm{H^\pm}$ & $tb$ & $W h_1$ & $W h_2$ & $W h_3$ \\
\hline
BP1 & 0.518 & $0.459$ & $2.26 \cdot 10^{-2}$ \\
BP2 & 0.840 & 0.105 & $5.38 \cdot 10^{-2}$ \\
BP3 & 0.133 & $2.31 \cdot 10^{-2}$ &
  $4.04 \cdot 10^{-3}$ & 0.839 \\
\end{tabular}\\[0.4em]
\begin{tabular}{l|ccc|ccc|ccc}
\textbf{Effective couplings} & $c_{h_1 VV} $ &
  $c_{h_1 t \bar t}$ & $c_{h_1 b \bar b}$
  & $c_{h_2 VV} $ &
  $c_{h_2 t \bar t}$ & $c_{h_2 b \bar b}$
  & $c_{h_3 VV} $ &
  $c_{h_3 t \bar t}$ & $c_{h_3 b \bar b}$\\
\hline
BP1 & -0.371 & -0.381 & 0.212 & 0.928 & 0.930 & 0.796 &
    $4.59 \cdot 10^{-2}$ & $-8.91 \cdot 10^{-2}$ & 7.40 \\
BP2 & 0.410 & 0.421 & 0.229 &
  -0.911 & -0.918 & -0.778 &
  $4.66 \cdot 10^{-2}$ & -0.192 & 4.23 \\
BP3 & 0.362 & 0.381 & 0.171 &
  -0.931 & -0.939 & -0.850 &
  $4.62 \cdot 10^{-2}$ & -0.272 & 3.18
\end{tabular}
\caption{\small 
\GWnn{Selection of benchmark points}
from the scan in \TBB{type~IV}
which describe the di-photon excess observed at
the LHC and the $b \bar b$ excess observed at LEP,
\TBB{while being in agreement with
all other experimental and theoretical constraints.}
\TBB{BP1 and BP3 additionally
\GWnn{predict a relic abundance of DM that is close to
the measured value},
whereas for BP2 the
predicted DM density is under-abundant.}
\label{tab:bp4}
}}
\end{table*}

\section{Conclusions and outlook}
\label{sec:conclu}

Recently, upon the inclusion of the
full Run~2 data set and improved 
analysis techniques, resulting in a
substantially improved sensitivity compared to the previous analysis,
the ATLAS collaboration has
reported an excess of $\sigATLAS\,\sigma$
local significance at about $\massATLAS \gev$
in the low-mass Higgs boson searches
in the di-photon final state.
An excess \SH{in the same channel} at 
\GW{the same}
mass value \TB{and with
higher local significance of~$2.9\, \sigma$
had previously been found by CMS
based on the Run~2 data set, 
\GWnn{and} an excess
with similar significance had also been observed
already in the \GWnn{CMS} di-photon searches at~$8\tev$.}
Neglecting possible 
correlations we obtain a combined signal strength of
$\muACgaga = \muAC^{+\dmuACp}_{-\dmuACm}\,,$
corresponding to an excess of $\sigAC\,\sigma$
\GW{for the mass value of}
$m_{\phi} \equiv m_{\phi}^{\rm ATLAS+CMS} = \massAC \gev\,$.

We have 
investigated the interpretation of 
\GW{the combined result from ATLAS and CMS}
as a 
di-photon resonance arising from the production of
a Higgs boson in the Two-Higgs doublet model that is extended by a
complex singlet (S2HDM).
\TBB{Using a genetic algorithm, we scanned the
parameter space of the model \GWnn{in order} 
to determine parameter
\GWnn{regions}
that feature a scalar state at~95~GeV
with sizable signal rates in the processes
in which the excesses have appeared.}
We have shown that a good description of the
excess that is in line with
\GW{the slightly increased significance of the combination in 
comparison to the previous result from CMS}
is possible in the Yukawa types~II and~IV, while being in agreement
with all other collider searches for additional
Higgs bosons, the measurements of the properties of the
SM-like Higgs boson at $125\gev$,
and further experimental and theoretical constraints. 
At the same time,
the model can account for
the observed 
DM relic abundance in agreement with the measurements
of the Planck satellite.

The di-photon excess observed at ATLAS and CMS
is especially intriguing in view of
additional excesses that appeared at
approximately the same mass.
Investigating this possibility, we have demonstrated that the S2HDM \GWnn{of} type~II can
simultaneously describe
the ATLAS/CMS di-photon excess and the $b \bar b$
excess observed at LEP, whereas no significant
signal for the CMS di-tau excess is possible 
in this model.
In the S2HDM \GWnn{of} type~IV, 
on the other hand, in addition
also a sizable signal 
strength in the di-tau channel can occur,
\TB{but with maximally reachable signal rates
somewhat below the signal strengths}
that would be
required to describe the
di-tau excess at the level of~$1\,\sigma$.

It should be noted
in this context
that our results in the S2HDM
can be generalised to other 
extended Higgs sectors containing 
\GWn{a SM-like Higgs boson and 
at least a second Higgs doublet as well as
at least one}
singlet with a Higgs boson at about $\massAC \gev$.

In the near future,
the \GW{possible} presence of a Higgs boson at about $\massAC \gev$
can be 
\GW{probed} by the eagerly awaited update of
the ATLAS searches in the di-tau final states covering the
mass region below $125\gev$.
\GW{Furthermore, the Run~3 results from ATLAS and CMS in the di-photon channel near $95 \gev$ will shed light on the question whether the 
excesses that have been observed by ATLAS and CMS in the di-photon channel have been a first sign
of a new particle.}
Further into the future, the existence of 
a possible state \hnf\ will
be tested in a twofold way at future Runs
of the (HL)-LHC, where the direct searches
for $\hnf$ \GW{in different channels} and the coupling measurements
of $\hotf$ will benefit 
in particular from a significant
increase of statistics.
However, it was demonstrated in \citere{Biekotter:2023jld}
that the experimental
precision of the coupling measurements of
the Higgs boson at $125\gev$ might not be
sufficient to exclude the S2HDM interpretation
of the excesses at $\massAC \gev$, or conversely 
confirm a deviation from the SM predictions.
On the other hand,
a future $e^+ e^-$ collider
could determine
the couplings of \hotf\ to a sufficiently high 
precision~\cite{Biekotter:2023jld}.
Despite the suppressed couplings of the
possible state at $\massAC \gev$ compared
to \hotf, a future $e^+e^-$ Higgs factory could produce
\hnf\ in large numbers if it has a
sufficiently large coupling to $Z$ bosons, see \GW{e.g.\ 
\citere{Drechsel:2018mgd},}
and determine its properties with high precision.

\medskip
In summary, the simultaneous observation of 
excesses in the $\gamma\gamma$ 
channel at the same mass value of
$\massAC \gev$ at both ATLAS
and CMS (together with the other observed excesses 
that are compatible with this mass value)
gives rise to the intriguing possibility that
a particle 
that cannot be accommodated by the SM of particle physics
could be discovered in the near future.

\section*{Acknowledgements}
We thank M.~Mart\'inez for helpful discussion on the
combination of the data.
S.H.~thanks the CTPU (Particle Theory and Cosmology Group)
at the IBS (Daejeon, South Corea) for
hospitality during the final stages of this work.
G.W.~acknowledges support by the Deutsche
Forschungsgemeinschaft (DFG, German Research
Foundation) under Germany‘s Excellence
Strategy -- EXC 2121 ``Quantum Universe'' --
390833306.
The work of G.W.~has been partially funded
by the Deutsche Forschungsgemeinschaft 
(DFG, German Research Foundation) - 491245950. 
S.H~acknowledges support from the grant IFT
Centro de Excelencia Severo Ochoa
CEX2020-001007-S funded by 
MCIN/AEI/10.13039/501100011033.
The work of S.H.~was supported in part by the
grant PID2019-110058GB-C21 funded by
MCIN/AEI/10.13039/501100011033 and by
``ERDF A way of making Europe''.
The work of T.B.~is supported by the German
Bundesministerium f\"ur Bildung und Forschung
(BMBF, Federal Ministry of Education and Research)
-- project 05H21VKCCA.

\bibliographystyle{JHEP}
\bibliography{refs}

\end{document}